\documentclass{emulateapj}
\usepackage{mathrsfs}
\usepackage{amssymb,amsmath,amsthm}
\newcommand\beq{\begin{equation}}
\newcommand\eeq{\end{equation}}
\newcommand\bea{\begin{eqnarray}}
\newcommand\eea{\end{eqnarray}}
\newcommand\bal{\begin{align}}
\newcommand\eal{\end{align}}

\newcommand\rmax{r_{{\rm max}}}

\renewcommand\bal{\mbox{\boldmath$\alpha$}}

\shorttitle{Soft X-ray excess of clusters and lensing }
\shortauthors{Lieu and Bonamente}

\begin{document}

\title{Soft X-ray excess of clusters: a thermal filament model, and
the strong lensing of background galaxy groups}

\author{Richard Lieu and Massimiliano Bonamente}
\affil{Department of Physics, University of Alabama, Huntsville, AL
35899.}

\begin{abstract}
The observational and theoretical status of the search for missing
cosmological baryons is summarized, with a discussion of some
indirect methods of detection.  The thermal interpretation of the
cluster soft X-ray and EUV excess phenomenon is examined in the
context of emission filaments, which are the higher density part of
the warm hot intergalactic medium (WHIM) residing at the outskirt of
clusters.  We derived an analytic radial profile  of the soft excess surface
brightness using a simple filament model, which provided us a means
of observationally constraining the WHIM parameters, especially the
total mass budget of warm gas associated with a cluster.  We then
pointed out a new scenario for soft excess emission, viz.  a
cluster that can strongly lens the soft X-rays from background WHIM
knots.  If, as seems quite likely, the missing baryons are mostly in
the WHIM halos of galaxy groups, the lensing probability will be
quite high ($\sim$ 10 \%).  This way of accounting for at least part
of a cluster's soft excess may also explain the absence of O VII
absorption at the redshift of the cluster.
\end{abstract}

\keywords{galaxies: clusters; cosmology: large-scale structure of universe}

\section{ Introduction}

The location of the `missing baryons' in the Universe is an open
and important question of contemporary cosmology, as interesting as
it is ironic, because the problem manifests itself as a deficit in
the mass budget which arises only at low redshifts, i.e. in the
space near us.  Observationally the total baryonic
content in stars, galaxies, and clusters of galaxies ($\Omega_b =
(2.1^{+2.0}_{-1.4})h_{0.7}^{-2}$ \%, Fukugita et al 1998) is only
about half of the amount required by Big Bang Nucleosynthesis models
($\Omega_b = (3.9 \pm 0.5) h_{0.7}^{-2}$ \%, Burles \& Tytler 1998)
or from measurements of the cosmic microwave background 
($\Omega_b = (4.6 \pm 0.2) h_{0.7}^{-2}$, Komatsu et al. 2008 WMAP5,
consistent with the Bennett et al 2003 WMAP1 and Spergel et al 2006 WMAP3 results).
%$\Omega_b = (4.4 \pm 0.4) h_{0.7}^{-2}$ \%, Bennett et al 2003 WMAP1, with small
%adjustments from WMAP3, Spergel et al 2006). 
Cosmological
hydrodynamic simulations have shown that this missing 50~\% 
of baryons is concealed in a tenuous filamentary gas of
temperature 10$^5$ - 10$^7$ K, currently referred by many to as the
WHIM (the Warm Hot Intergalactic Medium, Cen \& Ostriker 1999,
Dav\'e et al 2001).  It is also possible to derive this result
heuristically as follows.  Let $\lambda$ be a wavelength of bulk
motion of the intergalactic medium in the near Universe;  when
these waves collide and break, the thermal velocity of the shocked
gas will typically be $v \sim H_0 \lambda$, or $v \approx 100
h_{0.7} (\lambda/1.5~{\rm Mpc})$ km~s$^{-1}$ for $\lambda \sim$ the
cluster size.  Thus if gas heating takes place mainly at the `nodes'
of cluster-scale mass clumping, the thermal motion would involve a
value of $v$ that places the gas in the 10$^{5-7}$ K range of
temperatures.

Since the arrival of its theoretical prediction, search for the WHIM
has been an ongoing effort, with some success but no clinching
evidence as yet.  At the low end of the WHIM temperature scale, far
UV absorption lines have been reported (e.g., Richter
et al. 2008, Stocke et al. 2006, Tripp et al. 2006, Savage et al. 2005, Danforth and Shull 2005).
At temperatures where the bulk of the WHIM is expected to be, the
detection of O VII and O VIII absorption lines in the spectrum of
distant quasars (Nicastro 2005) is a more debatable result (Kaastra
et al 2006). Although soft X-ray and EUV emission in regions of
galaxy concentration (Werner et al 2008, Mannucci et al 2007,
Zappacosta et al 2002, Mittaz et al 1998) may also be the signature
of warm filaments, the definitive proof of this interpretation, viz.
an identifiable line at the appropriate redshift, is still not
available.  More precisely, the clinching line signature of O VII
was claimed (Kaastra et al 2003, Finoguenov et al 2003) and refuted
(Lieu \& Mittaz 2005, Takei 2008).

We examine if limits can be placed on the WHIM from less
direct measurements.  
%The column density of free electrons in the
%intergalactic medium is a parameter to which value the WHIM can
%contribute the same independently of its state of clumping.  
The column density of free electrons in the intergalactic medium
is proportional to the average WHIM density $\rho_{{\rm WHIM}}$
in a given volume, and it is independent of its state of clumping.
To see this, allow the WHIM matter to reside in clumps of number density
$n$, each having mass density $\rho$ and radius $r$.  By mass
conservation inside a large spherical volume of radius R, we have
$\rho_{{\rm WHIM}} \sim n\rho r^3$.  The average electron column
along a random direction and for a given WHIM ionization fraction,
being proportional to the product of the mass column $\rho r$ of one
clump and the number of intercepted clumps $\sim nr^2 R$, will then
$\sim \rho_{{\rm WHIM}} R$,  i.e. a constant for any sightline of
length $\sim R$ that penetrated our large volume.

There are several ways of constraining this electron column.
The Sunyaev-Zel'dovich effect (of the WHIM) will be at a very low
level, as will reionization, with $\delta T_{{\rm CMB}}/T_{{\rm
CMB}} \sim \sigma_{{\rm Th}} n_e \ell kT/(m_e c^2) \sim 10^{-7}$ for
characteristic WHIM parameters at the minimum (homogeneous)
overdensity of $\delta = \Omega_{{\rm WHIM}}/\Omega_c \approx$ 0.02,
viz. 
\beq \begin{cases}
kT \approx 0.1~{\rm keV}~(T~\approx~10^6~{\rm K});\\
~n_e
\approx 10^{-7} \left(\frac{\delta}{0.02}\right)
\left(\frac{h}{0.7}\right)^2~{\rm cm}^{-3};\\
~\ell \approx 2~{\rm
Gpc}, 
\end{cases}
\eeq 
where in obtaining $n_e$ we assumed a fully ionized pure
hydrogen plasma.
For comparison, the Sunyaev-Zel'dovich effect of a typical
massive cluster ($T=10^8$K, $n_e=10^{-3}-10^{-4}$ cm$^{-3}$, $l\sim 2$ Mpc)
is $\delta T_{{\rm CMB}}/T_{{\rm
CMB}} \sim 10^{-4}-10^{-5}$ (see Hern\'{a}ndez-Monteagudo et al. 2008
for a discussion of the SZE from the WHIM).
%(at 10$^6$ K the ionization fraction is $>$ 99 \%,
%Mazzotta et al 1998).   
Another limit might be afforded by the
realization that such a column can cause frequency dependent delay
on the timescale of minutes to hours in the arrival of $\sim$ 100
MHz emission from distant quasars, except quasars with so rapid a
variability as to avail themselves for this test are those very ones
affected by plasma scintillations in our local interstellar medium
(Dennett-Thorpe and de Bruyn 2002). Angular broadening of quasars
caused by WHIM-like scintillation was assessed by Lazio et al 2008,
who concluded that the only way of securing a useful observational
limit is if an AGN is found to `twinkle' at a position close to that
of a pulsar, as data about the latter will enable us to take out the
interstellar effects of our Galaxy.  Thus, the situation regarding
these `tangential' probes is that they too do not deliver any useful
verdict.

Here we take a step backwards by returning to the prospect of direct
WHIM filament detection at the outskirts of clusters, for reasons
that would soon become clear.  The most powerful argument for
excluding any warm gas association with the central soft X-ray
excess of clusters (e.g. Fabian 1996) is the large radiative cooling
rate, since the gas (which at 10$^6$ K is already at the peak
temperature of its cooling curve) has to be clumped to co-exist with
the hot cluster medium, making it radiate even faster. On the other
hand, the scenario of soft photons seen in projection from a
line-of-sight WHIM filament at the cluster's outskirts is more
attractive, not only because the lack of physical contact between
the two phases now alleviates the warm gas from its former problems,
but also because numerical codes of structure formation (Cen \&
Ostriker 1999, Dav\'e 2001, Cheng et al 2005) do expect such
filaments to preferentially converge at clusters and groups, which
are the `knots' of the WHIM network.

The purpose of this paper is to calculate the first analytical
formula for the radial profile of soft X-ray surface brightness by
employing a simple filament model.  We shall see that this already
affords a means of placing rather useful observational constraints
on the WHIM parameters at the outskirt of a cluster, including and
especially the total mass of the warm baryons.  We will then propose
the possibility of a cluster's soft excess being due in part to the
strong lensing of background WHIM clumps, the largest population of
which is galaxy groups.  This is quite a radical approach to the
problem:  it may explain the lack of identifiable O VII absorption
at the redshift of the candidate cluster, because if distant WHIM
emissions are superposed and focused by the cluster then their O VII
absorptions would be likewise, leading to a smearing of the line, or
even a `re-location' of the line to an altogether different redshift.

\section{ Difficulties with a straightforward WHIM
interpretation of the cluster soft X-ray excess}

We examine more carefully why the idea mentioned in the previous
section of the central soft excess being emitted by warm
intracluster gas unrelated to the WHIM was not well received. Since
the center of a cluster is permeated by the hot X-ray gas at (or
close to) the virial temperature, the warm component can only exist
for any conceivable length of time if it is clumped into dense
clouds to ensure pressure equipartition between the two phases.  Now
the typical parameters of the hot gas are $kT \approx$ a few keV and
$n_e \sim$ 10$^{-3}$ cm$^{-3}$, and because the pressure requirement
implies equality of the products $n_e T$, we have $n_e \gtrsim$ 0.01
cm$^{-3}$ for these clouds.  The immediate problem is the radiative
cooling time, \beq \tau_{{\rm warm}} = 6 \times 10^8
\left(\frac{T}{10^6~K}\right)^{\frac{1}{2}}
\left(\frac{n_e}{0.01~{\rm cm}^{-3}}\right)^{-1}~~{\rm years}, \eeq
which means the cloud cools and collapses on a timescale far shorter
than the age of a cluster: it is thermally unstable (note that in
eq. (2) we omitted the contribution from line emission which will
reduce $\tau_{{\rm warm}}$ even more).  Apart from cooling, the
cloud has difficulty in sustaining itself against photo-ionization
by the hot gas, which occurs in a time $\tau_{{\rm photo}} \approx 2
\times 10^7/(F/10^4~{\rm ph~cm}^{-2}~{\rm s}^{-1})~{\rm years}$ for
O VII, where $F$ is the X-ray flux from the gas.  Moreover, any
`balance' that may result from the two opposing mechanisms of
cooling and ionization is bound to be extremely precarious. 
% Why
%should equilibrium be found at the most unstable position, viz. the
%peak of the cooling curve?

In Section 1 we also presented the cluster's soft X-ray excess
emission as possible evidence for the WHIM.  
Among the possible models for the soft excess (e.g., see review by
Durret et al. 2008) is the Cheng et al. (2005) proposal of very dense gas
in pressure equilibrium associated with merging structures within clusters.
This model relieves the cooling problem described above in this section,
since its non-equilibrium  emission can be sustained for periods longer than
those calculated in Equation 2.
Alternatively, the WHIM may be seen in projection against the cluster's
X-ray emission, although the filaments' brightness is approximately one order
of magnitude too faint
to explain the soft excess flux (e.g., Dolag et al. 2006; Mittaz et al. 2004).
To date, the soft excess has been detected in several clusters at low redshift;
the Bonamente et al. (2002) blind survey of the excess emission found that $\sim 50$\%
of the clusters have statistically significant evidence for the excess. Since clusters
with high S/N observations preferentially have soft excess, the previous
statistic is probably a lower limit.
 The excess flux is typically 
$\sim 10-20$\% relative to the hot ICM contribution in the soft X-ray band,
with a trend of increasing relative flux at larger radii (e.g., Lieu et al. 1999).

 Let us now elaborate
on the filament origin of the soft excess, 
by adopting as representative WHIM parameters those
of the thermal model of Coma cluster's soft excess, as listed under
`warm component 1' of Lieu et al 1996, Table 3 (see also Tables 2
and 3 of Bonamente et al 2003)~\footnote{According to Lieu et al. (1996),
the warm component in the 0--3 and 9-12 arcmin regions 
has a spectral normalization constant of $3.7\pm1.0 \times 10^{-5}$ and
$1.2\pm0.6 \times 10^{-5}$, respectively; this normalization is proportional to the
EM via $\text{EM} = 2.3 \times 10^{68} \times Norm \times \text{(area of
annulus in arc minutes square).}$}.  
We select two regions,  the 0 -- 3
arcmin and 9 -- 12 arcmin annuli, to investigate if the radial
profile of soft X-ray surface brightness can accommodate the
filament model.   For these regions the temperature is still $kT
\sim$ 0.1 keV, but it is the emission integral of the optically thin
gas filament, with the value 
\beq {\rm EI} = {\rm EM} \cdot {\cal A}
= n_e^2 L^* {\cal A} = 4.79 \times 10^{65}~{\rm cm}^{-3}, 
\label{eq:ei}
\eeq 
for 0
-- 3 arcmin, where $L^*$ and ${\cal A}$ are respectively the
effective length and cross-sectional area of the emitting column,
and EM is the emission measure (or column density of emission), that
has the physical significance. For a redshift of $z =$ 0.0232 and a
Hubble constant of $h=$ 0.7, we have ${\cal A}=$ 2.15 $\times$
10$^{47}$ cm$^2$. Hence the effective length of the filaments (i.e.
the total length of sections of emitting plasma (that may belong to
different filaments) intercepted by the line of sight) is 
\beq L^* =
n_e^{-2} \cdot{\rm EM} \approx \left(\frac{n_e}{10^{-3}~{\rm
cm}^{-3}}\right)^{-2}~{\rm Mpc}. \eeq 
Moving outwards to the 9 -- 12
arcmin annulus, $L^*$ is found to fall by 3.5 times. 
The observed EI rises by 2.3 times from the value in Equation \ref{eq:ei},
viz. EI $=$ 1.10 $\times$~10$^{66}$ cm$^{-3}$, causing $L^*$ to 
decrease by $\sim$ three times because the projected area ${\cal A}$ 
of this annulus is seven times larger than that of  the 0 -- 3 arcmin 
circle.
  Beware that
if the electron density of $n_e =$ 10$^{-3}$ cm$^{-3}$, which
corresponds to a rather large WHIM overdensity of $\delta \approx$
200, is lowered, $L^*$ will easily become $\gg$ 1 Mpc, i.e. the
filaments will have a dimension far exceeding the cluster scale,
placing them at `intercluster' venues which further makes it harder
to justify the use of $\delta\approx$ 200.

\section{Analytic radial profile of the soft X-ray surface
brightness of WHIM filaments}

%There is ample awareness among WHIM investigators of the `promise'
%held by the filament model, viz. that linear emitters protruding
The filament model, viz. linear emitters protruding
outwards from beyond some cluster radius, can account for the
radially rising trend of the soft X-ray excess reported by
observations (Durret et al. 2008; Bonamente et al 2002, 2001; Lieu et al 1999).  Yet the
surface brightness profile of such a configuration was never
evaluated even under the simplest scenario to see if this `promise'
is deliverable.   We attempt to do so here, by invoking a model of
$N$ filaments, each of length $L$, cross sectional area ${\cal A}$,
and electron density $n_e$, which converge to cover a fraction $f_0$
(by area) of the spherical surface at some radius $R$ - the radius
of filamentary `footpoints' (see Figure 1).   The WHIM filling factor at any radius
$r$ is then given by 
\begin{equation} f(r) = 
\begin{cases}
\frac{N {\cal A}}{4 \pi r^2} = f_0 \left(\frac{R}{r}\right)^2 \hspace{1cm} \text{for $r \geq R$}\\
0 \hspace{1cm} \text{ for $r<R$}. 
\end{cases}
\end{equation}
The soft X-ray surface brightness  of the WHIM, for a line
of sight with impact parameter $b$, is then given by $S_X(b) = {\rm
EM} \cdot {\Lambda_{ee}}/{4 \pi}$, in which $\Lambda_{ee}$ is the
WHIM plasma emissivity, and the emission measure 
\beq {\rm EM} =
n_e^2 L^* = n_e^2 \int_{-\infty}^{\infty} f(r) dz = 2 f_0 n_e^2 R^2
\int_{\tilde r}^{\infty} \frac{dr}{r\sqrt{r^2 - b^2}}, 
\eeq 
where
$\tilde r = R$ for $b < R$ and $\tilde r = b$ for $b \geq R$.

\begin{figure}
\vspace{0cm}
\begin{center}
\centering
\includegraphics[angle=0,width=2.5in]{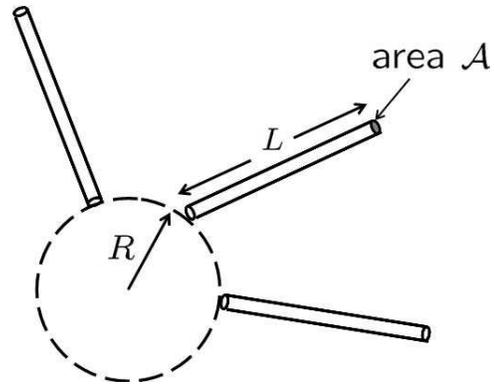}
\vspace{0cm}
\end{center}
\caption{A simple model of WHIM filaments converging to some
critical radius $R$ of a cluster; at this radius the surface cover
factor by the filaments reaches the maximum value of $f_0$ mentioned
in the text.  Each filament has length $L$, cross sectional area
${\cal A}$, and is optically thin to soft X-rays.}
\end{figure}

The integral of eq. (6) may readily be evaluated, yielding 
\beq {\rm
EM} = 
\begin{cases}
\frac{2 f_0 n_e^2 R^2}{b} \tan^{-1} \left(\frac{b}{\sqrt{R^2 -
b^2}}\right),~{\rm for}~ b < R;\\
\frac{f_0 n_e^2
\pi R^2}{b},~{\rm for}~b \geq R.
\end{cases}
 \eeq 
The two results match each
other smoothly at the `footpoint' radius $b=R$ (as they should), and
this is also the radius of maximum brightness.  Beyond $b=R$, ${\rm
EM}$ falls off with $b$ as ${\rm EM}\sim 1/b \sim 1/\theta$ where
$\theta$ is the angular distance away from the cluster center.
Beneath $b=R$, however, EM is almost a constant, changing
(decreasing) only slightly from ${\rm EM}= \pi f_0 n_e^2 R$ at $b=R$
to ${\rm EM}=2f_0 n_e^2 R$ at $b=0$. Since we understood from
section 2 why the footpoints have to be at $R \gtrsim$ a few Mpc to
ensure `segregation' between the warm and hot gases, and since all
the observations of the rising radial trend of soft excesses (see
references above) have thus far involved physical radii less than a
few Mpc, this means it is the $b<R$ behavior of EM that matters. The
encouraging news is that here, the near flatness of the profile over
the entire scale height of the virialized X-ray cluster emission can
account for the observed trend of the relative soft excess.

In order to compare this model with the Bonamente et. al (2003)
measurements of the soft X-ray surface brightness in the
neighborhood of the Coma cluster, we plot in Figure 2 the surface
brightness $S_X$ for the parameter values $R=1$ Mpc, $n_e=10^{-3}$
cm$^{-3}$, $f_0=0.5$, and with the emissivity  being
that of a plasma at $kT=0.1$ keV, metal abundance $A=0$, and
averaged over the ROSAT 1/4 keV band using the APEC model of Smith
et al. (2001) at $\Lambda_{ee}=4.5 \times 10^{-16}$ 
counts cm$^3$ s$^{-1}$. The red
line in Figure 2 represents the 1/4 keV ROSAT All-Sky Survey
background within 2-5 degrees of Coma. Given that the excess
emission at the outskirts of Coma is on par with the local 1/4 keV
background, we see that the filament model can account for the
observed magnitude of the outer soft excess.

\begin{figure}
\vspace{0cm}
\begin{center}
\centering
\includegraphics[angle=-90,width=3.0in]{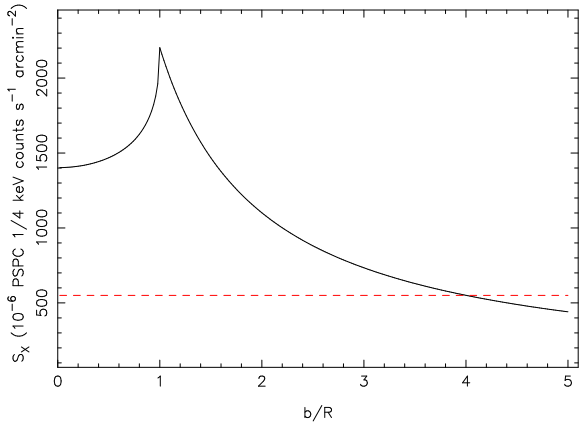}
\vspace{0cm}
\end{center}
\caption{Intensity of the filament emission following the model of
Section 3. The surface brightness has been multiplied by the average
effective area $A_{eff} \simeq 150$ cm$^2$ of the PSPC instrument in
the 1/4 keV band (R2 band, Snowden et al. 1998), in order to compare
this detector-dependent intensity to the value measured by Bonamente
et al. (2003) in the neighborhood of the Coma cluster (shown as the
red dashed line). }
\end{figure}

Such a simple model yields no central peaking of the absolute
surface brightness, which could be a consequence of radial
dependence in $n_e$, ${\cal A}$, or $L$; although in this paper we
would like to discuss an altogether different mechanism that might
also play an important role.  Note that the effective emission
length is $L^* \sim f_0 R$ irrespective of the actual filament
length $L$. The parameter $L$ does determine the total mass of the
WHIM associated with the cluster, which is
\begin{eqnarray}
M_{{\rm WHIM}} &=& \int_R^L 4\pi r^2 n_e m_p f(r) dr 
=4.25 \times 10^{14} \nonumber\\
& &\left(\frac{n_e}{10^{-3}~{\rm
cm}^{-3}}\right)^{-3} \left(\frac{f_0}{0.5}\right)^{-1}
\left(\frac{L}{5~{\rm Mpc}}\right)\nonumber\\
& & \left(\frac{{\rm EM}}{3 \times
10^{18}~{\rm cm}^{-5}}\right)^2~{\rm M}_\odot 
\end{eqnarray}
Eq. (8) constrains the filament model: there is far less room for
manipulation of the parameters $n_e$ and $L$ than previously thought
(EM is fixed by observations), and arbitrary choice of their values
can easily lead to an unacceptable proliferation of the mass budget.

\section{The mass profile of a cluster of galaxies}

Before discussing the way in which lensing by a cluster's
gravitational field can affect its soft excess emission, we first
summarize our understanding of the mass distribution of clusters,
which is found to follow the generalized NFW profile (see Mahhavi et
al 2007 for a review). Although in the `authentic' version of it
(Navarro et al 1995, 1996, 1997) the density $\rho(r) \sim 1/r$ for
small $r$, steepening gradually to $1/r^2$ at intermediate $r$ and
finally reaching $1/r^3$ at large $r$, the points at which these
transitions occur are not sharp (i.e. the index of $r$ changes
gradually and continuously), and are determined by the core radius
and concentration parameter, both of which can vary from cluster to
cluster.  
%Especially for the inner radii, there is considerable
%debate about the form of $\rho (r)$, it ranges from the advocates of
%a flatter than a $1/r$ dependence (Tonini et al 2006) to the highly
%cited cold dark matter simulation work of Moore et al 1998 which
%concluded upon an inner slope of $r^{-\alpha}$ where $|\alpha|
%\gtrsim$ 1.4.  

Especially for the inner radii, there is considerable
debate about the form of $\rho (r)$. For example, Tonini et al. (2006) 
and Schmidt and Allen (2008) 
find  flatter than a $r^{-1}$ dependence, Navarro et al. (2004)
and Diemand et al. (2004, 2005) find inner slopes 
consistent with the $1/r$ slope 
($r^{-\alpha}$ with $\alpha \simeq 1.1\pm0.4$), while
Moore et al. (1998) 
determined an inner slope of $\alpha \gtrsim$ 1.4.
In any case, even those who favor a more gentle slope
of $|\alpha| \lesssim$ 1 invoked high concentration parameter that
indicates considerable mass inside these relatively smaller radii.
Moreover, little is known about the {\it nuclear} region of $r
\lesssim$ 10 kpc, where the frequent presence of a bright central
galaxy could cause a re-steepening of the slope.  
%A common feature
%which most authors seem to find is the $\rho (r) \sim 1/r^2$ form of
%the singular isothermal sphere (SIS), from $r \approx$ 20 kpc to $r
%\approx$ 300 kpc 

We shall adopt the best-fit NFW profile of the combined Subaru and
Hubble ACS data of the cluster Abell~1689 (Broadhurst et al 2005, Diego
et al 2005), which represent a comprehensive, model independent weak
and strong lensing mass survey of the cluster, as a sufficiently
representative mass profile for the purpose of this paper.  From
Table 3 of Broadhurst et al 2005 one finds the SIS density scaling
of $\rho (r) \sim 1/r^2$ between approximately $r=r_{{\rm min}}=$
27$/h_{0.7}$ kpc and $r=r_{{\rm max}}=$ 430$/h_{0.7}$ kpc. At the
$r=$ (14 -- 27)$/h_{0.7}$ kpc radii the slope has still not reached
$|\alpha| =$ 1, and further inwards are {\it terra incognita}: as
already mentioned the value of $|\alpha|$ here could increase again.
The normalization for $\rho (r)$ within the SIS zone corresponds to
an enclosed mass per unit radius of \beq \frac{M(r)}{r} = 3 \times
10^{15}~{\rm M}_\odot~{\rm Mpc}^{-1},~{\rm for}~r_{{\rm min}} \leq r
\leq r_{{\rm max}}, \eeq irrespective of the value of the Hubble
constant $h$.

 The choice of Abell~1689 as a representative case is
motivated  by the following considerations. Abell~1689 is a very relaxed massive cluster, according
to the morphology of its X-ray emission (e.g., Xue and Wu 2008), with no evidence
of significant substructure;
moreover, the mass profile based on the lensig data of Broadhurst et al. (2005) is
smooth in the region of interest, lending further support to its relaxed nature;
finally, the lensing data for this cluster is of the highest
quality available.
%According to the ROSAT analysis of Bonamente et al. (2002; the only available
%for Abell~1689), this cluster has no strong evidence for soft excess emission. 

\section{Gravitational lensing of soft X-ray emission from
background WHIM clouds}

In setting up the prerequisites let $b$ be the impact parameter of a
light ray under the influence of some spherically symmetric cluster
lens, and $b_0$ that of the same ray in the absence of the lens.
Since the lens does not expand with the Hubble flow, $b$ and $b_0$
are physical dimensions independent of the epoch of lensing.
Provided $b$ falls within the SIS part of the cluster mass profile,
i.e. $b$ satisfies $r_{{\rm min}} \leq b \leq r_{{\rm max}}$ where
the two limits were defined in section 4,  the deflection angle
$\psi$ will be given by \beq \psi = \frac{4GM}{c^2
\rmax}\left[\arccos\left(\frac{b}{\rmax}\right)+
\frac{\rmax-\sqrt{\rmax^2- b^2}}{b}\right], \eeq  (see e.g. Lieu \&
Mittaz 2005).  Essentially $\psi$ is a constant within this SIS
zone: it ranges from $\psi=2\pi GM/(c^2\rmax)$ at $r \ll \rmax$ to
$\psi= 4GM/(c^2 \rmax)$ at $r \lesssim \rmax$. Adopting the former,
and applying eq. (9), one obtains $\psi = 9.3 \times 10^{-4}$ which
we shall henceforth approximate as \beq \psi = \psi_0 \approx
10^{-3}. \eeq 
The lens equation 
\beq
\psi (b) = \frac{b-b_0}{D}
\eeq 
then simplifies to
\beq b_0 = b-D\psi_0, \eeq
where we defined
\beq D=\frac{D_l D_{ls}}{D_s}, \eeq
and the reader is referred to Figure 3 for the meaning of the
various comoving distance indicators on the right side of eq. (13).

\begin{figure}
\vspace{0cm}
\begin{center}
\centering
\includegraphics[angle=0,width=3in]{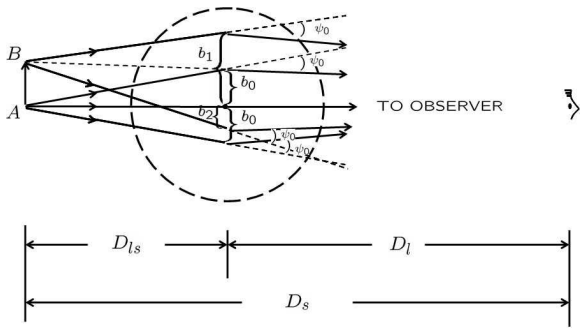}
\vspace{0cm}
\end{center}
\caption{Strong gravitational lensing of a background source AB by a
singular isothermal sphere (SIS) which deflects light by a constant
angle $\psi_0$.  The two rays from point A that reach the observer
after they are bent by the lens (not drawn to scale) have the impact
parameters $b= \pm b_0$, and cast images of A that form part of an
axially symmetric ring - the Einstein ring of physical radius
$r_E=|b_0|$ on the lensing plane. The impact parameter of the
unlensed ray that connects point B to the observer O also happens to
be $b=b_0$. After lensing, the actual rays that allow O to see B
have impact parameters of opposite signs, viz. $b=b_1$ and $b=b_2$,
where $|b_1| = 2|b_0|= 2r_E$. If the arrow AB were to rotate
azimuthally about the optically axis AO to form a circular source
disk of unlensed radius $|b_0|=r_E$ on the lensing plane, the
resulting image will also be a circular disk, but of radius $2r_E$
on this same plane.  Hence the brightness magnification $m$ of the
source will be four.  If AB has smaller size than currently, then
$m>$ 4. If AB is larger, we will have $m<$ 4 and the source will be
too big to fit within the strong lensing limit.}
\end{figure}

Now consider a small element of the background source that maps to
an element on the lensing plane defined by radial and transverse
intervals $db$ and $b\,d\varphi$, with an area of
$dA=bdb\,d\varphi$.  When the lens is removed, the rays from this
same source element would pass through another element with area
$dA_0=b_0\,db_0\,d\varphi$. The area (or brightness) magnification
of this element due to lensing is therefore
 \begin{equation}
m=\left|\frac{dA}{dA_0}\right|=
\left|\frac{b\,db}{b_0\,db_0}\right|.
 \end{equation}
The modulus signs are needed because the ratio can be negative if
caustics are involved, leading to image reversals.

We proceed to identify the region of strong lensing, which has the
characteristic scale length $r_c$ given by the circular caustic that
are images of the central spot, where the magnification is infinite.
This limiting radius, known as the Einstein radius $r_E$, is fixed
by the condition $b_0=0$ which becomes, after applying eq. (12),
\beq r_E = D\psi_0. \eeq  For finite $b$ whilst maintaining
$b\approx r_E$, we have from eq. (12) \beq b_0 = b- D\psi_0 = b-r_E.
\eeq For $b \gtrsim r_E$ the right side of eq. (15) is positive. For
$b \lesssim r_E$, however, it is negative, and the source pixel is
on the other side of the optical center as the image pixel, with
values of $\varphi$ differing by $\pi$; images here are reversed. In
this case we can still write \beq |b_0| = r_E -b = r_E - |b|. \eeq
Thus for every source pixel with $|b_0|<r_E$ there are \emph{two}
image circles, with radii $|b|=r_E \pm |b_0|$.  As shown in Figure
3, a small and centrally aligned source disk of radius $|b_0|=r_E$
is imaged twice, once onto the annulus between $|b|=r_E$ and
$|b|=2r_E$ and once, reversed, onto the disk $|b|<r_E$.  For this
source, it's clear that $|db|=|db_0|$, so the magnifications of the
two images are, from eq. (14), \beq
m_{\pm}=\left|\frac{b}{b_0}\right|=\frac{r_c}{|b_0|}\pm 1. \eeq The
total magnification is $m = m(b_0) = m_+ (b_0) + m_- (b_0) =
2r_E/|b_0|$. Hence a centrally aligned source of projected physical
radius $|b_0| = r_E$, or area $A_0 = \pi r_E^2$ on the lensing plane
{\it and without} the lens, will become four times brighter with the
help of the lens, because it will then occupy the new area of \beq A
= \int_0^{r_E} 2\pi m(b_0) b_0 db_0 = 4\pi r_E^2. \eeq  Further,
according to eq. (18) a smaller aligned source will be magnified by
even more, while a much larger source will not be affected very much
at all, i.e. it will only be weakly lensed.

There are two limiting scenarios of relevance to WHIM emissions.  If
a WHIM source lies closely behind the cluster because it is a
filament section visible to the observer, we will have $D_l \approx
D_s$, $r_E \approx D_{ls} \psi_0$, and the physical size of the
source will also be $\approx b_0$.  Hence the strong lensing
criterion of $b_0 \lesssim r_E$ would imply a filament width $w
\lesssim D_{ls} \psi_0$. Since $D_{ls} \lesssim L$ where $L \approx$
5 Mpc is the scale length of a filament, we obtain after applying
eq. (11) the requirement of $w \lesssim$ 5 kpc.   It would seem
unlikely that WHIM filaments are as thin as this.  Perhaps an even
more serious problem is the smallness of the impact parameters $b$
for rays from such a source, as $b \sim r_E \approx$ 5 kpc also,
i.e. now the lensing takes place inside the {\it nuclear} part of
the cluster where, as already discussed in section 4, the density
profile is not well known and is probably $\rho (r) \sim 1/r$, which
gives rise to {\it less} lensing than that of the SIS profile
currently being considered.  In any case, the effect of this kind of
brightening process is spatially very limited, it will at best
enhance the soft excess in the central few kpc of the cluster which
is not easily resolved by X-ray observatories.

The more interesting second possibility is one that involves distant
background WHIM clouds, which {\it could} find themselves lying
directly behind our optical center because as will be shown below
the scenario under which any random sightline may intercept one of
these clouds with appreciable chance is not so demanding.  Assuming
(reasonably) that the WHIM network extends to a distance $\ell\sim$
1 Gpc behind the cluster and the observer is much closer to the
cluster than that, we now have $D_{ls} \approx D_s \gg D_l$, $r_E
\ll D_{ls} \psi_0$, but the strong lensing criterion of $b_0
\lesssim r_E$ still constrains the physical radius of the source to
$a \lesssim D_{ls} \psi_0$. For a `mid-way' source of $D_{ls} \sim$
0.5$h_{0.7}^{-1}$ Gpc this yields $a \lesssim$ 0.5$h_{0.7}^{-1}$ Mpc
with the help of eq. (11). The lensing is then `self consistent'
because it falls within the SIS zone of the cluster, and moreover
soft X-ray emitting WHIM clouds with scale sizes 0.5 Mpc or smaller
are found by both simulations (see Figure 7 of Mittaz et al 2004a)
and observations (see Soltan et al 1996, 1997 and Finoguenov et al
2007, especially the last paper) to be associated with galaxy groups
and galaxy halos.

In fact, if all of the $n \approx$ 1.56 $\times$ 10$^{-4} h_{0.7}^3$
Mpc$^{-3}$  population of galaxy groups of the ESO survey (Ramella
et al 2002) possesses soft X-ray halos of 0.5 Mpc radius and at a
density of $n_e \approx$ 10$^{-3}$ cm$^{-3}$, this would be
sufficient to account for all the missing baryons of section 1 as
WHIM constituents.  The halo interception probability  by a
sightline through the cluster center (or any sightline) is 
\begin{eqnarray} p =
\pi na^2\ell &=& 12 \left(\frac{n}{1.56 \times 10^{-4}~{\rm
Mpc}^{-3}}\right) \nonumber\\
 & & \left(\frac{a}{0.5~{\rm Mpc}}\right)^2
\left(\frac{\ell}{1~{\rm Gpc}}\right) \% 
\end{eqnarray}
for $h=$0.7,  and is
by no means negligible as already mentioned.  The product of $p$ and
the emission measure of EM~$\approx n_e^2 a$ per halo gives the
average EM for the sightline in the same form as it was used it
sections 2 and 3.  It is 
\begin{eqnarray} {\rm EM} &=& 2 \times 10^{17}
\left(\frac{n}{1.56 \times 10^{-4}~{\rm Mpc}^{-3}}\right)
\left(\frac{n_e}{10^{-3}~{\rm cm}^{-3}}\right)^2 \nonumber\\
 & & \left(\frac{a}{0.5~{\rm Mpc}}\right)^3 \left(\frac{\ell}{1~{\rm
Gpc}}\right)~{\rm cm}^{-5} 
\end{eqnarray} for $h=$ 0.7.  Without strong
lensing, EM is as given by eq. (21) and is $\approx$ 2.5 times below
even the brightness of Coma's 9 -- 12 arcmin soft excess, section 2.
It is therefore barely distinguishable from the background.  With
lensing, however, EM is enhanced by a factor $\approx$ 4, eq. (19),
and can now account for 50 \% of the 0 -- 3 arcmin soft excess.

Moreover, the lensed flux is superposed over a cluster radius of
$r_{{\rm image}} = 2r_E =$ where $r_E$ is defined in eqs. (15) and
(13).  For the value of $D_{ls} =$ 0.5$h_{0.7}^{-1}$ Mpc we adopted,
and using $D_l =$ 100$h_{0.7}^{-1}$ Mpc as Coma's distance, we
obtain $r_{{\rm image}} \approx$ 0.17$h_{0.7}^{-1}$ Mpc assuming a
flat Universe where $D_s = D_l + D_{ls}$, whereas the mean physical
radius of Coma's 9 -- 12 arcmin annulus is 0.31$h_{0.7}^{-1}$ Mpc.
Thus the flux is {\it not} dispersed into a large area that includes
the outskirts; rather, it affects a substantial and resolvable part
of the cluster's core region.  This can then represent a
non-negligible fraction of the soft excess there.  Note that the
calculation here is conservative, because each background WHIM halo
was treated as uniform.  If their emissions are centrally peaked
(due e.g. to a radial scaling of the filaments'  density, cross
sectional area, or length, section 3), the strong lensing will
enhance the surface brightness of these peak regions by more than a
factor of four, eq. (18), and the resulting observed brightness
profile of soft X-rays will likewise be even sharper and more
interesting.  Under such a scenario, lensing of background halos
could account for {\it all} of the inner soft excess of some
clusters.

Given the degrees of freedom of our present model and
the uncertainty in the mass profile gradient, it is not possible to 
establish with certainty what fraction of the excess is due to strong lensing.
Our estimate of a 10\% probability of lensing, together with the
detection of soft excess in $\geq$50\% of the clusters, suggests that 
additional mechanisms are responsible for the excess, especially at large radii.
Thermal emission due to dense gas (Cheng et al. 2005) or by the filaments
themselves (e.g., Mittaz et al. 2004; Dolag et al. 2006), or non-thermal
emission (e.g., Sarazin and Lieu 1998; Blasi and Colafrancesco 1999) are viable
alternatives to complement the contribution due to the lensing effect.\rm 

\section{Conclusion}

Although there are various indirect ways of detecting the WHIM and
confirming (or refuting) the hypothesis that most of the missing
baryons at low redshift are in this phase, the more promising search
would still appear to be those involving emission and absorption of
soft X-rays in the vicinity of clusters of galaxies where the WHIM
is expected to congregate.

In this paper we developed a simple model of WHIM filaments to
account for the surface brightness profile of the outer cluster
soft X-ray excess, in particular the radial rise of the fractional
excess relative to the normal X-ray emission.  The inner excess,
however, necessitates a separate intepretation.  We propose that
this is due to the strong lensing of background soft X-ray halos
(most likely associated with galaxy groups) by the central gravitational
potential of clusters.  When applied to the Coma cluster, using 
reasonable parameters for these halos, we have the prospect of
explaining $\sim$ 50 \% of the 0 -- 3 arcmin soft excess
The calculations in this paper invokes filaments of constant length, 
cross-sectional area
and electron density, and thus they must be considered as a proof-of-principle
for a more realistic model; for example,
one which allows the filament
properties to change with cluster radius.  In the case of some
clusters, a possible additional contribution to this excess is the
strong lensing by the cluster of WHIM emissions associated with
background galaxy groups.  If a central gradient is present in the
the emission of some of these aligned sources, the lensing will be
particularly effective, and will further modify the soft excess
profile of the candidate cluster.

\section*{References}

\noindent Bahcall, N., 1998, ARA\&A, 26, 631.

\noindent Blasi, P. and Colafrancesco, S. 1999, APh 12, 169

\noindent Bonamente, M., Lieu, R., \& Mittaz, J.P.D., 2001, ApJ,
547, L7.

\noindent Bonamente, M., Lieu, R., Joy, M., \& Nevalainen, J., 2002,
ApJ, 576, 688.

\noindent Bonamente, M., Joy, M., \& Lieu, R., 2003, ApJ, 585, 722.

\noindent Broadhurst, T., Takada, M., Umetsu, K., Kong, X., Arimoto,
N., Chiba, M., \& \\
\indent Futamase, T., 2005, ApJ, 619, L143.

\noindent Cheng, L. -M. et al, 2005, A \& A, 431, 405.

\noindent   Danforth, C., Shull, M., 2005, ApJ, 624, 555.

\noindent Dennett-Thorpe, J., \& de Bruyn, A.G., 2002, Nature, 415,
57.

\noindent Diego, J. M., Sandvik, H. B., Protopapas, P., Tegmark, M.,
Benítez, N., \& \\
\indent Broadhurst, T., 2005, MNRAS, 362, 1247.

\noindent Diemand, J., Moore, B. and Stadel, J. 2004, MNRAS, 353, 624

\noindent Diemand, J., Zemp, M., Moore, B., Stadel, J and Carollo, M. 2005, MNRAS, 364, 665

\noindent {Dolag}, K. and {Meneghetti}, M. and {Moscardini}, L. and {Rasia}, E. and 
	{Bonaldi}, A., 2006, MNRAS,  370, 656

\noindent {Durret}, F., {Kaastra}, J.~S., {Nevalainen}, J., {Ohashi}, T., \& {Werner}, N.
  2008, Space Science Reviews, 134, 51

\noindent Finoguenov, A., et al, 2007, ApJS, 172, 182.

\noindent Hern\'{a}ndez-Monteagudo, C.,  Trac, H.,  Verde, L. and Jimenez, Raul 2008, ApJ 652, 1.
\noindent Kaastra, J., et al, 2006, ApJ, 652, 189.

\noindent Komatsu, E. et al. 2008, arXiv:0803.0547v2

\noindent Lazio, T.J.W., 2008, ApJ, 672, 115.

\noindent  Lieu, R., Bonamente, M., Mittaz, J.P.D., Durret, F., Dos
Santos, S., \& \\
\indent Kaastra, J.S., 1999, ApJ, 527, L77.

\noindent Lieu, R., \& Mittaz, J.P.D., 2004, ASSL, 309, 155
(astro-ph/0409661).

\noindent Lieu, R., \& Mittaz, J.P.D., 2005, The Identification of
Dark Matter, p18,\\
\indent World Scientific, Singapore (astro-ph/0501007).

\noindent Lieu, R., \& Mittaz, J.P.D., 2005, ApJ, 628, 583.

\noindent Mahdavi, A., Hoekstra, H., Babul, A.,  Sievers, J., Myers,
S. T., \& \\
\indent Henry, J. P., 2007, ApJ, 664, 162.

\noindent Mannucci, F., Bonnoli, G., Zappacosta, L., Maiolino, R.,
\& Pedani, M., \\
\indent 2007, A \& A, 468, 807.

\noindent Mazzotta, P., Mazzitelli, G., Colafrancesco, S., \&
Vittorio, N., 1998, A \& AS, 133, 403.

\noindent Mittaz, J.P.D., Lieu, R., \& Lockman, F.J., 1998, ApJ,
498, L17.

\noindent Mittaz, J.P.D., Lieu, R., \& Cen, R., 2004a, ASSL, 309,
171 (astro-ph/0409661).

\noindent Mittaz, J., Lieu, R., Cen, R., \& Bonamente, M., 2004b,
ApJ, 617, 860.

\noindent Moore, B., Governato, F., Quinn, T., Stadel, J., \& Lake,
G., 1998, ApJ, 499, L5.

\noindent Navarro, J.F., Frenk, C.S., \& White, S.D.M. 1995, MNRAS,
275, 56.

\noindent Navarro, J.F., Frenk, C.S., \& White, S.D.M. 1996, ApJ,
462, 563.

\noindent Navarro, J.F., Frenk, C.S., \& White, S.D.M. 1997, ApJ,
490, 493.

\noindent Navarro, J.F. et al. 2004, MNRAS, 349, 1039

\noindent Ramella, M., Geller, M.J., Pisani, A., \& da Costa, L.N.,
2002, AJ, 123, 2976.

\noindent   Richter, P., Paerels, F. B. S. and  Kaastra, J. S. 2008,
Space Science Reviews, 134, 25.

\noindent   Savage, B., Lehner, N., Wakker, B., Sembach, K., Tripp, T.,
2005, ApJ, 626, 776.

\noindent Schmidt, R.W. and Allen, S.W. 2007, MNRAS, 379, 209.

\noindent Smith, R., Brickhouse, Nancy S., Liedahl, D. and
 Raymond, J., 2001, ApJ, 556, L91.

\noindent Snowden, S.,  Egger, R., Finkbeiner, D.,  Freyberg, M.,
 Plucinsky,  P., 1998 , ApJ, 493, 715

\noindent Soltan, A. M., Hasinger, G., Egger, R., Snowden, S., \&
Truemper, J., 1997, A \& A, 320, 705.

\noindent Soltan, A. M., Hasinger, G., Egger, R., Snowden, S., \&
Truemper, J., 1996, A \& A, 305, 17.

\noindent Stocke, J., Penton, S., Danforth, C., Shull, M., Tumlinson, J., McLin, K.
2006, ApJ, 641, 217.

\noindent Tonini, C., Lapi, A., \& Salucci, P., 2006, ApJ, 649, 591.

\noindent   Tripp, T., Aracil, B., Bowen, D., Jenkins, E.,
2006, ApJ, 643, 77.

\noindent Xue, S.-J. and Wu, X.-P. 2008, ApJ, 576, 152

\noindent Werner, N., Finoguenov, A., Kaastra, J.S., Simionescu, A.,
Dietrich, J.P., \\
\indent Vink, J., \& Böhringer, H., 2008, A \& A, n482, L29.

\noindent Zappacosta, L. et al. 2002, A \& A, 394, 7.

%Zappacosta, L., Maiolino, R., Mannucci, F., Gilli, R., \&
%Schuecker,  P., \\
%\indent 2005, MNRAS, 357, 929.

\end{document}